\documentclass[preprint]{elsarticle}
\usepackage{lineno,hyperref}
\usepackage{natbib}
\usepackage{multirow}
\usepackage{color}

\definecolor{DarkBlue}{rgb}{0.00,0.00,0.55}
\hypersetup{
 colorlinks=true,
 citecolor=DarkBlue,
 linkcolor=DarkBlue,
 urlcolor=DarkBlue}

\modulolinenumbers[5]

\journal{Journal of \LaTeX\ Templates}

\def\arcsec{\hbox{$^{\prime\prime}$}}
\def\utw{\smash{\rlap{\lower5pt\hbox{$\sim$}}}}
\def\udtw{\smash{\rlap{\lower6pt\hbox{$\approx$}}}}

\def\farcs{\hbox{$.\!\!^{\prime\prime}$}}

\biboptions{authoryear}

\begin{document}

\begin{frontmatter}

\title{Small-scale magnetic and velocity inhomogeneities in a sunspot light bridge}

\author{Rohan E. Louis\corref{mycorrespondingauthor}}
\cortext[mycorrespondingauthor]{\color{blue}rlouis@aip.de}
\address{Leibniz-Institut f\"ur Astrophysik Potsdam (AIP),
	  An der Sternwarte 16, 14482 Potsdam, Germany}

\begin{abstract}
High resolution spectro-polarimetric observations of a sunspot light 
bridge by {\em Hinode}, reveal small-scale inhomogeneities in the 
magnetic field and velocity. These inhomogeneities arise as a 
consequence of a weak, secondary lobe in the Stokes $V$ profile which 
have a polarity opposite that of the sunspot and very large 
($>$5~km~s$^{-1}$) Doppler velocities of both signs, suggesting
two distinct types of magnetic anomalies. These two sets of 
inhomogeneities are highly time-dependent and appear exclusively 
in the upper half of the light bridge and only after the light 
bridge is completely formed. Both sets of inhomogeneities appear as 
patches and can be present independent of the other, next to one 
another, or spatially separated in a single scan. A two-component 
inversion of the corresponding spectral profiles indicate that the 
inhomogeneities occupy a very small fraction, amounting to less 
than 10\%, of the resolution element. These structures 
are likely driven by small-scale magneto-convection
where they could further interact with the overlying sunspot magnetic
field to produce reconnection jets in the chromosphere.
\end{abstract}

\begin{keyword}
Sunspots, light bridges, magnetic fields, high resolution, photosphere, spectro-polarimetry \\
Received 8 June 2015 / Accepted 2 September 2015
\end{keyword}

\end{frontmatter}


\section{Introduction}
\label{intro}
Light bridges (LBs) are conspicuous, bright structures in the 
umbrae of sunspots and pores and are typically present during the
formation or fragmentation of spots \citep{1987SoPh..112...49G}. 
LBs can be regarded as either field-free intrusions 
of hot plasma in the gappy umbral magnetic field 
\citep{1979ApJ...234..333P,1986ApJ...302..809C}, or 
manifestations of large-scale magneto-convection
\citep{2008ApJ...672..684R} in umbrae. The latter 
scenario is now widely, although not universally, accepted
as being responsible for the fine structure and energy 
transport in sunspots. This in particular has been made possible 
from the advancement in numerical simulations
\citep{2006ApJ...641L..73S,2010ApJ...720..233C}, as well as high resolution
observations from both the ground and space
\citep{2010ApJ...713.1282O,2010ApJ...718L..78R,2013A&A...549L...4R,
2013A&A...553A..63S,2015ApJ...803...93E}.
The interaction between umbral dots and intruding penumbral filaments,
typically seen during light bridge formation 
\citep{2007PASJ...59S.577K,2012ApJ...755...16L}, lends further 
credence to magneto-convection in sunspots. In this sense, LBs 
represent a natural location where convective disruptions are more
vigorous and apparent than in other parts of a sunspot. Such 
a disruption could produce magnetic inhomogeneities 
that might explain the pronounced chromospheric activity in LBs as 
reported in \citet{2008SoPh..252...43L} and recently in 
\citet{2014A&A...567A..96L}. In this paper I analyse 
small-scale magnetic anomalies in a sunspot LB using
high resolution observations from {\em Hinode}
\citep{2007SoPh..243....3K}.

\section{Observations}
\label{obs}
High resolution observations of NOAA AR 11271
were acquired by the {\em Hinode} spectropolarimeter 
\citep[SP;][]{2008SoPh..249..233I,2013SoPh..283..579L}
on 2011 August 18--19. During these two days the SP took 14
scans (nine scans on August 18 and five scans on August 19)
of the active region in the fast mode, each scan 
covering a field of view (FOV) of nearly 75\arcsec$\times$82\arcsec.
On August 18, the scans were taken at 11:00--11:16~UT,
11:20--11:36~UT, 11:40--11:56~UT, 
18:12--18:28~UT, 18:32--18:48~UT,
18:52--19:08~UT, 19:12--19:28~UT,
19:32--19:48~UT, and 19:52--20:08~UT.
The five SP scans on August 19 have a similar scan time of 16~min
starting at 08:05~UT and ending at 10:21~UT. In the 
fast mode, the SP recorded the four
Stokes profiles of the Fe~{\sc i} lines at 630~nm with a 
spectral sampling of 2.15~pm, a step width of 0\farcs29, and 
a spatial sampling of 0\farcs32 along the slit. The Level 0 
data were reduced using standard 
routines included in the Solar-Soft package \citep{2013SoPh..283..601L}.
The active region traversed heliocentric angles of 
42$^\circ$ to 30$^\circ$ between August 18 and 19.

\begin{figure}
\centerline{
\includegraphics[angle=90,width = \textwidth,clip=]{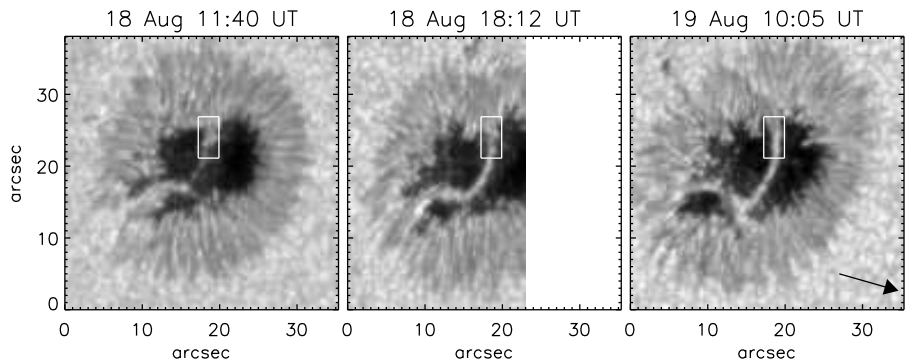}
}
\vspace{-20pt}
\caption{Continuum images of the leading sunspot in AR 11271 on 2011 
August 18 and 19. The white rectangle depicts a 10$\times$19~pixel 
FOV shown in Fig~\ref{fig02}. The arrow in the right panel points 
to disk centre.}
\label{fig01}
\end{figure}

\section{Results}
\label{res}

\subsection{Anomalous Stokes profiles in the light bridge}
\label{ano}
Figure~\ref{fig01} shows a small LB in the south-eastern half of 
the leading sunspot in AR 11271 that forms during the early part 
of August 18. There is also an indication of a pronounced
intrusion of penumbral filaments at the northern part of the 
umbra-penumbra boundary (white rectangle in Fig~\ref{fig01}). The 
tip of these filaments extend into the umbra as diffuse umbral dots 
(left panel of Fig~\ref{fig01}). Nearly 6.5~hr later a second LB 
forms, dividing the umbra into two nearly equal halves with one 
its ends coinciding with the intruding penumbral filaments 
described above, while its other end connects to the smaller 
southern LB. The larger LB in the sunspot is the object of 
interest for this paper. There are frequent transitions 
between formation and fragmentation over a period of 
3~days, but the LB remains intact between the latter 
half of August 18 to the end of August 19 (Fig.~8 of 
\cite{2014A&A...570A..92L}).

\begin{figure}
\centerline{
\includegraphics[angle=90,width = 1.1\textwidth,clip=]{stokes-profiles-selected-1.eps}
}
\vspace{-10pt}
\centerline{
\includegraphics[angle=90,width = 1.1\textwidth,clip=]{stokes-profiles-selected-2.eps}
}
\vspace{-10pt}
\caption{Stokes $V$ profiles in the LB. The FOV corresponds 
to the white rectangle shown in Fig.\ref{fig01}. The black 
contour outlines the LB. The $V$ profile corresponding to 
the Fe line at 630.25~nm is overlaid on each pixel and is 
scaled between $\pm$10\% of the QS continuum intensity. The 
blue (red) colours indicate that the weaker, second lobe is 
present on the blue-ward (red-ward) side of the primary lobe. 
The time indicated on the top of the panel represents the 
instant when the slit was at the left edge of the FOV.}
\label{fig02}
\end{figure}

A smaller FOV consisting of the upper part of the LB is shown in 
Fig.~\ref{fig02}, wherein the Stokes $V$ profile at every pixel 
has been overlaid on the continuum map. Prior to the formation 
of the LB, the profiles consist of normal, anti-symmetric lobes
(top left panel of Fig~\ref{fig02}). However, after the LB is 
formed, there is a clear indication of two different kinds of 
anomalous profiles present in it. These are indicated in blue 
and red colours in Fig.~\ref{fig02} and will henceforth be 
referred to as blue profiles (BPs) and red profiles (RPs). 
These anomalous profiles comprise a weak, but discernible, 
second component and their name serves to distinguish their 
position with respect to the primary lobes of the $V$ profile. 
It is to be noted that, although both sets of profiles have 
opposite Doppler (velocity) signatures, their signs are opposite 
to the primary lobes of the $V$ profile. This would suggest 
offhand that the BPs and RPs have a polarity opposite that 
of the sunspot (see Sect.~\ref{sir}).

The {\color{black}amplitude of the secondary lobe in Stokes $V$,
corresponding to the} weak component in the BPs and RPs, is less than 
5\% {\color{black}(e.g. bottom panels of 
columns 1 and 2 in Fig.\ref{fig03})}.  
While the profiles 
indicated in the figure were selected by hand, it was verified 
that the amplitude of the weak lobe was at least four times 
the noise level of 1.5$\times 10^{-3}$. The peak corresponding 
to the BPs and RPs is, on an average, located at 17~pm and 22.5~pm, 
respectively from line centre. This suggests that the features 
are associated with very large doppler velocities. It is observed 
that the BPs and RPs are confined exclusively to the upper half 
of the LB throughout the observing period. These profiles appear 
in patches, sometimes as small as 0.2~arcsec$^2$ in area. The 
BPs and RPs can either appear i) completely independent of 
each other (panels 2 and 3), or ii) adjacent to each other 
(panels 4 and 5), or iii) separated from each other (panel 6). 
Since there are only limited SP scans it is difficult to establish 
the lifetimes of these features. As they can be 
sometimes seen in successive scans at/near the same pixel, a 
lower limit of 20~min can be attributed to these features.

\subsection{Magnetic field and LOS velocity associated with BPs and RPs}
\label{sir}
\begin{figure}
\centerline{
\includegraphics[angle=0,width = 0.3\textwidth,clip=]{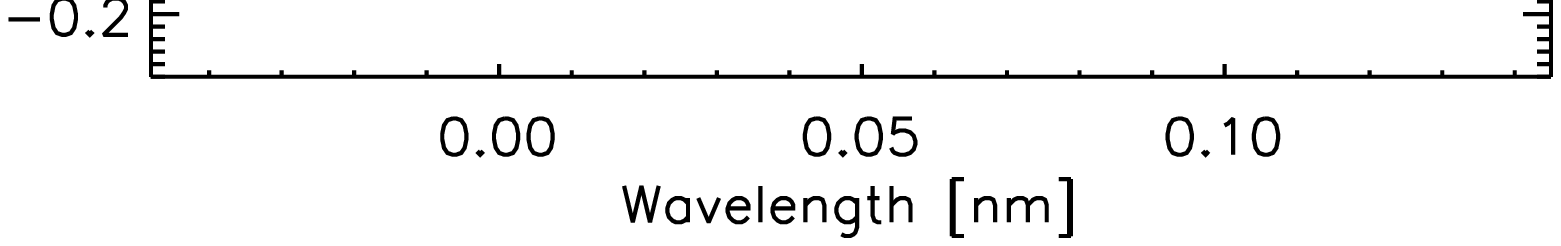}
\includegraphics[angle=0,width = 0.3\textwidth,clip=]{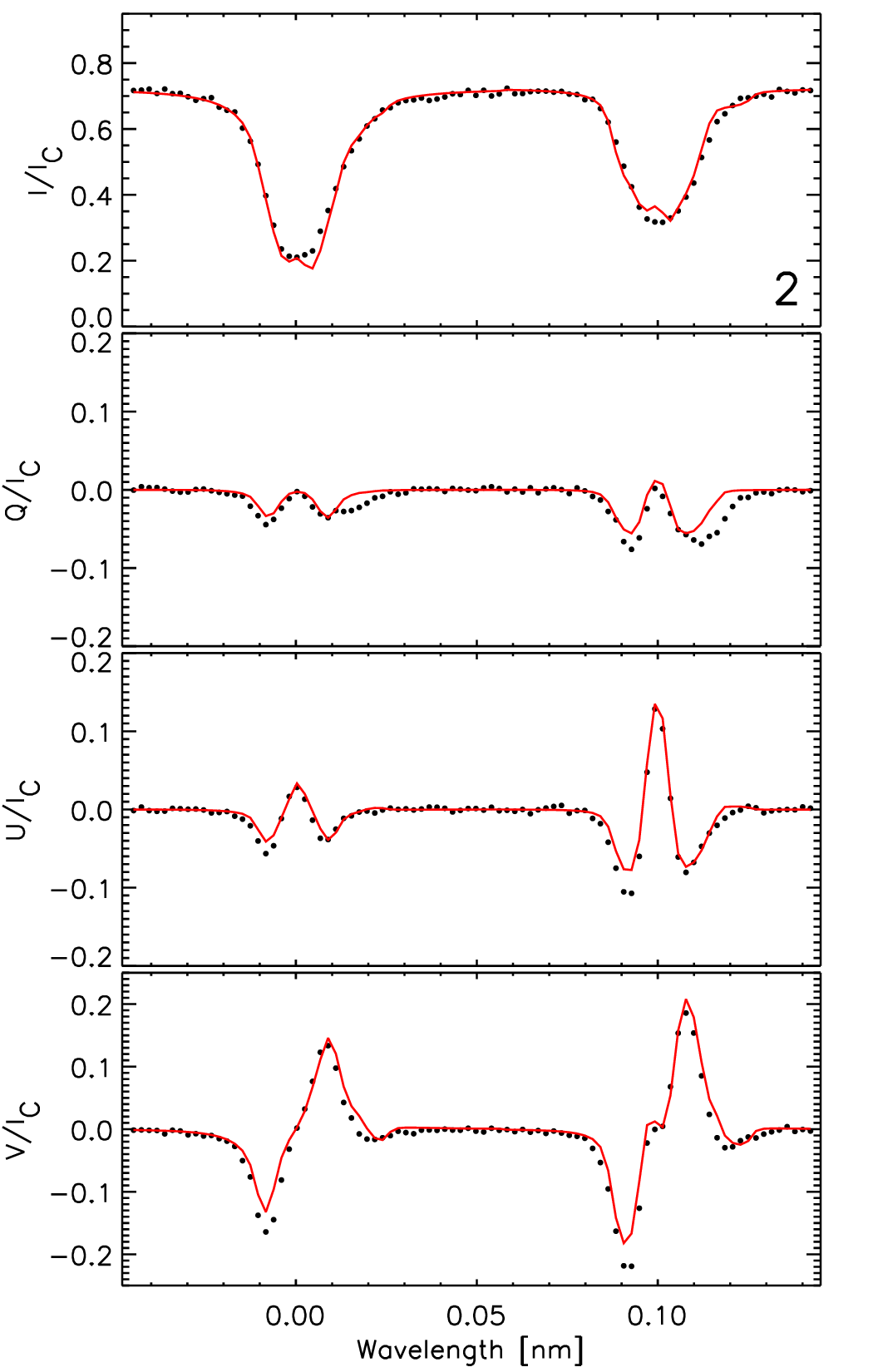}
\includegraphics[angle=0,width = 0.3\textwidth,clip=]{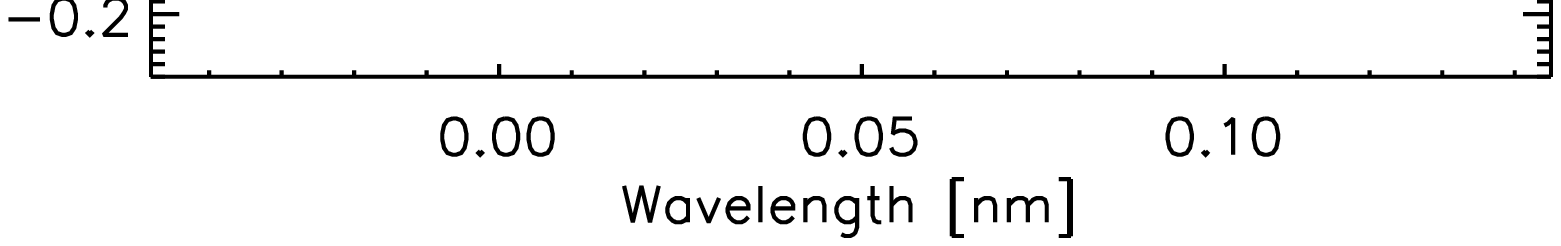}
\includegraphics[angle=0,width = 0.3\textwidth,clip=]{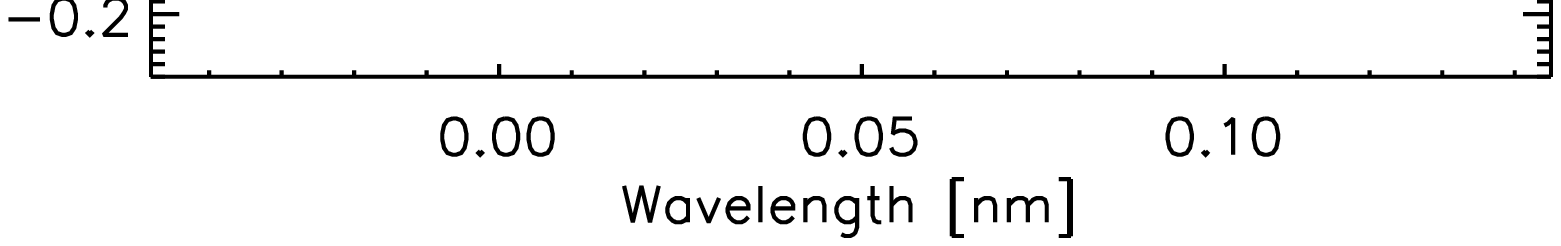}
}
\vspace{-5pt}
\caption{Results from SIR inversions. The {\em filled black circles} and 
{\em solid red line} represent the observed and synthetic spectra, 
respectively. From left to right: Stokes profiles corresponding to BP, RP,
neighbouring pixel in LB, and an umbral pixel.}
\label{fig03}
\end{figure}

The BPs and RPs signify multiple components in the resolution 
element or gradients in the magnetic field and LOS velocity 
\citep{1978A&A....64...67A}. The physical parameters associated 
with these profiles were retrieved using Stokes Inversion 
based on Response functions \citep[SIR;][]{1992ApJ...398..375R}.
A typical BP and RP were inverted using a two-component model 
with height independent parameters except for temperature, which 
was perturbed with two nodes. Results of this inversion are shown 
in columns 1 and 2 of Fig.~\ref{fig03} which correspond to the BP 
and RP, respectively. Columns 3 and 4 represent fits to a neighbouring 
pixel in the light bridge and an umbral pixel, respectively, which were
obtained using a single component inversion with height-independent
parameters. Attempts were made with a single component 
inversion with gradients in the physical parameters but the resulting 
fits were unsatisfactory. Additionally, the two-component inversion 
was redone with height-dependent parameters, as described in 
\citet{2013A&A...549L...4R}, but there was only a marginal improvement 
in the fit. The height-independent case allows a more simplistic 
interpretation of the physical scenario while still reproducing the
spectral signatures specific to the anomalous profiles.

\begin{table}[!h]
\begin{center}
\caption{Values of physical parameters obtained from SIR inversions. Field 
inclination is expressed in the local reference frame.}
\begin{tabular}{l|cccc}
\hline
  &    BP  &   RP   &   LB  & Umb \\
 \hline\hline
\multirow{2}{*}{Field Strength [G]}         &   1554 &   1581 &   1688 &   2611\\
                                            &   1535 &   1747 &      0 &      0\\
\hline
\multirow{2}{*}{Field Inclination [deg]}    &    150 &    145 &    156 &    164\\
                                            &     59 &     54 &      0 &      0\\
\hline
\multirow{2}{*}{LOS Velocity [km~s$^{-1}$]} &    0.6 &    0.3 &    0.2 &    0.3\\
                                            &   -5.2 &    6.7 &    0.0 &    0.0\\
\hline
\multirow{2}{*}{Fill Fraction}              &   0.91 &   0.96 &   1.00 &   1.00\\
                                            &   0.09 &   0.04 &   0.00 &   0.00\\
\hline
\end{tabular}
\label{tab01}
\end{center}
\end{table}

Table~\ref{tab01} shows that the second component, corresponding to 
the BP and RP, is characterised by Doppler velocities that are sub-sonic, 
but in excess of 5~km~s$^{-1}$, and a polarity opposite that of the 
sunspot.  A majority of the pixels in the LB which show normal Stokes 
$V$ profiles also have the same polarity as the sunspot. In addition, 
one finds that the fill fraction of the second component to be around 
10\%, which represents the upper limit for these structures and
the reason for the weak lobe in the profile. The field strength of 
the second component exceeds 1.5~kG and is comparable with the 
primary/main component which is also similar to other pixels in the LB. 

\subsection{Discussion}
\label{discuss}
The anomalous profiles in the LB represent small-scale 
inhomogeneities in the magnetic field and velocity. These 
inhomogeneities have a polarity opposite that of the 
sunspot, including the LB, and have large Doppler-shifts
of either sign. Their presence is detected by the weak, 
second component in the circular polarization profiles, 
which is clearly discernible, despite i) the reduced 
spatial resolution of the SP fast mode scan, and ii) not 
having corrected for spatially-induced polarized stray 
light arising from the point spread function of the 
{\em Hinode} telescope \citep{2012A&A...548A...5V}. The 
reason these inhomogeneities stand out is because of their 
large Doppler shifts, the absence of which would have 
rendered them difficult to detect because of their 
small fill fraction, as they constitute less than 10\%
of the resolution element.

The anomalous Stokes profiles in a sunspot LB, reported by 
\citet{2009ApJ...704L..29L}, had the same polarity 
as the sunspot and were primarily associated with supersonic 
downflows, while those reported here have a polarity opposite 
that of the sunspot and have large Doppler shifts of either 
sign, that appear very close to one another and evolve on short 
time-scales. {\color{black}The red profiles in particular, are 
similar to those associated with the supersonic Evershed mass 
flux \citep{1909MNRAS..69..454E} returning to the solar photosphere 
in the outer penumbra and beyond the sunspot boundary
\citep{2001ApJ...547.1130W,2001ApJ...547.1148W,2001ApJ...549L.139D,
2004A&A...427..319B,2007PASJ...59S.593I,2008A&A...480..825B}.}

An inspection of broad-band {\em Hinode} 
Ca~{\sc ii}~H filtergrams on August 18 reveal that small-scale 
chromospheric jets, seen the following day \citep{2014A&A...567A..96L}, 
are only present after the LB is formed, i.e. after 18:00~UT, but not 
earlier in the day when the LB was only confined to an extended 
penumbral structure. Since the supersonic downflows detected by 
\citet{2009ApJ...704L..29L} were also associated with strong 
chromospheric enhancements/jets, it strongly suggests a causal 
relationship between the small-scale inhomogeneities 
in a LB, irrespective of their magnetic polarity and nature of 
the Doppler shift, and the dynamics in the overlying chromosphere. 

The natural question that arises is, what produces these small-scale 
inhomogeneities. Considering that these features have large Doppler-shifts,
a strong gas pressure gradient is needed to drive the plasma to these 
velocities. Furthermore, the pressure gradient ought to have opposite 
signs in order to produce both blue and redshifts. {\color{black}
It is known that in the inner penumbra, penumbral grains exhibit an 
inward motion towards the umbra-penumbra boundary, which is observed both
in intensity images as well as LOS velocity maps \citep{2009ApJ...694.1080S}. 
This inward migration could explain the non-uniform, but predominantly 
unidirectional, flow directed from the northern end of the LB into its axis, 
as found by \citet{2014A&A...567A..96L}. Such flows would indicate
the existence of a pressure gradient that could drive a flow along the axis
of the LB, at least at the northern end or upper half of the LB, where the
small-scale velocity and magnetic inhomogeneities are predominantly located. 
However, it is also known that penumbral grains are associated with blueshifts of 
around 1~km~s$^{-1}$ in the limb side penumbra \citep{2006ApJ...646..593R}. 
Thus, their motion and weakly subsonic speeds, leaves the presence of the 
strong redshifts in the LB, unexplained.} 

Recently \citet{2014A&A...568A..60L} reported 
supersonic downflows at the edges of light bridge granules and 
suggested that these downflows reflect intense convection that 
arises from a combination of gravitational acceleration and radiative 
cooling at the interface of the umbra and the LB. However, the 
LB in question here has not evolved sufficiently to a granular 
stage and is still confined to the umbral environment, which is why
the small-scale chromospheric jets are prevalent here and absent in 
the LB studied by \citet{2014A&A...568A..60L}. Furthermore, 
the redshifts reported here occur at different locations of the LB 
and often in its interior which would suggest that the magnetic 
and velocity inhomogeneities described here are likely produced 
by small-scale magneto-convection which would effectively produce both 
blue and redshifts. This is supported by the fact that the G-band images
indicate the presence of a dark lane running along the axis of the LB
(Figs. 1 and 2 of \citet{2014A&A...567A..96L}), even
though the resolution is insufficient to detect smaller granule/grain-like 
structures on the LB. However, the photospheric morphology of the LB 
investigated here is similar to the non-granular LB studied by 
\citet{2010ApJ...718L..78R}, with the exception that they only 
detect upflows of around 1~km~s$^{-1}$, along the dark lane, that 
are surrounded by relatively weaker downflows. Small-scale magneto-convection 
would also explain the opposite polarity magnetic elements in the LB
which could reconnect with the overlying the sunspot magnetic field 
to produce the chromospheric jets. This scenario is also consistent
with spectroscopic observations of \citet{2003SoPh..215..261S}, where
they found rapid variations of blue and redshifts extending over a 
large section of a LB. These variations, measuring nearly 1.5~km~s$^{-1}$, 
occurred over a time scale of 35~s. 

If small-scale magneto-convection is indeed responsible for these
inhomogeneities it is necessary to determine how the physical 
parameters are stratified in geometrical height 
\citep{2010ApJ...720.1417P} and how the iso-$\tau$ layers are 
arranged across the LB. Unfortunately, the significant height 
difference in the formation heights of the LB and the neighbouring 
umbra makes the estimation of such a geometrical height scale, solely 
from the photospheric Fe line pair, impossible and would necessitate 
the use of multiple spectral lines that form over a wide range of 
heights in the solar atmosphere. This would be possible with, for 
instance, the GREGOR Fabry-P{\'e}rot Interferometer 
\citep[GFPI;][]{2012AN....333..880P}, where high resolution
imaging spectro-polarimetric observations in two spectral lines 
can be performed quasi-simultaneously with sufficiently high 
spectral resolution.

\subsection{Conclusions}
\label{concl}
The magnetic and velocity inhomogeneities in a
sunspot light bridge are likely produced by magneto-convection
that renders high-speed plasma flows at very small-spatial scales
that also vary in time. The presence of these inhomogeneities 
could facilitate magnetic reconnection in the chromosphere with 
the overlying sunspot magnetic field and illustrates the complex 
magnetic topology associated with light bridges. Multi-wavelength, 
spectro-polarimetric observations with good spectral and high 
spatial resolution are necessary to confirm the magneto-convective 
origin of these small-scale inhomogeneities.
 
\begin{acknowledgements}
{\footnotesize
Hinode is a Japanese mission developed and launched by
ISAS/JAXA, collaborating with NAOJ as a domestic partner 
and NASA and STFC (UK) as international partners. Scientific 
operation of the Hinode mission is conducted by the Hinode 
science team organized at ISAS/JAXA. This team mainly consists 
of scientists from institutes in the partner countries. Support 
for the postlaunch operation is provided by JAXA and NAOJ (Japan), 
STFC (UK), NASA, ESA, and NSC (Norway). I thank Horst Balthasar 
for reading the manuscript and providing useful comments and 
suggestions. I am grateful for the financial assistance from 
the German Science Foundation (DFG) under grant DE 787/3-1 and 
from SOLARNET--the European Commission's FP7 Capacities Programme 
under Grant Agreement number 312495. I thank the referees for their
useful comments and suggestions.
}
\end{acknowledgements}

\bibliography{louis_reference}

\end{document}